# SUANPAN: Scalable Photonic Linear Vector Machine

Ziyue Yang[1,5], Chen Li[1,5], Yuqia Ran[2], Yongzhuo Li[1]✉, Xue Feng[1]✉, Kaiyu Cui[1], Fang Liu[1], Hao Sun[1], Wei Zhang[1], Yu Ye[2], Fei Qiao[1], Cun-Zheng Ning[3], Jiaxing Wang[4], Connie J.Chang-Hasnain[4], Yidong Huang[1]✉

[1]Department of Electronic Engineering, Tsinghua University, 100084 Beijing, China.
[2]State Key Laboratory for Mesoscopic Physics and Frontiers Science Center for Nano-Optoelectronics, School of Physics, Peking University, 100871 Beijing, China.
[3]College of Integrated Circuits and Optoelectronic Chips, Shenzhen Technology University, 518118 Shenzhen, China.
[4]Berxel Photonics Company Ltd., 518071 Shenzhen, China.
[5]These authors contributed equally: Ziyue Yang, Chen Li.
✉e-mail: liyongzhuo@tsinghua.edu.cn; x-feng@tsinghua.edu.cn; yidonghuang@tsinghua.edu.cn

**Abstract**

Photonic linear operation is a promising approach to handle the extensive vector multiplications in artificial intelligence techniques due to the natural bosonic parallelism and high-speed information transmission of photonics. Although it is believed that maximizing the interaction of the light beams is necessary to fully utilize the parallelism and tremendous efforts have been made in past decades, the achieved dimensionality of vector-matrix multiplication is very limited due to the difficulty of scaling up a tightly interconnected or highly coupled optical system. Additionally, there is still a lack of a universal photonic computing architecture that can be readily merged with existing computing system to meet the computing power demand of AI techniques. Here, we propose a programmable and reconfigurable photonic linear vector machine to perform only the inner product of two vectors, formed by a series of independent basic computing units, while each unit is just one pair of light-emitter and photodetector. The element values of one vector are encoded in the time domain, *i.e.* the output duration of continuous light-emitter while those of the other are encoded in the space domain, *i.e.* the emitter-detector position. The result of the inner product is obtained by the sum of photocurrents of all photodetectors. Since there is no interaction among light beams inside, extreme scalability could be achieved by simply duplicating the independent basic computing unit while there is no requirement of large-scale analog-to-digital converter and digital-to-analog converter arrays. Our architecture is inspired by the traditional Chinese Suanpan or abacus and thus is denoted as photonic SUANPAN. As a proof of principle, SUANPAN architecture is implemented with an 8×8 vertical cavity surface emission laser array and an 8×8 $MoTe_2$ two-dimensional material photodetector array. The experimental computing fidelities for randomly generated vector inner products are all over 98% for 1-bit, 2-bit, 4-bit and 8-bit quantization and over 95% for 8-80 vector dimensionalities with 4-bit quantization, respectively. Two typical AI tasks of the Ising machine for NP-hard optimization problem and artificial neural network for visual perception are performed to demonstrate the ability of SUANPAN architecture. For the Ising problem, 1024-dimensional problems are successfully solved, which is the highest dimensional optical Ising machine with heuristic algorithm. For artificial neural network, a competitive classification accuracy of 84%~88% is achieved for MNIST (Modified National Institute of Standards and Technology) handwritten digit dataset. We believe that our proposed photonic SUANPAN is capable of serving as a fundamental linear vector machine that can be readily merged with existing electronic digital computing system and is potential to enhance the computing power for future various AI applications.

**Introduction**

Artificial intelligence (AI) is currently an active topic in both scientific research and commercial application as well as daily life[1,2]. For AI techniques, the linear operations of high-dimensional vectors are fundamental and dominant in both the artificial neural networks[3-5] (ANN) and optimization problem solvers, *such as the* Ising machine[6-8]. As the complexity of problems increases, the dimensionality of the processed vector grows rapidly, resulting in a huge computational burden. It is known that vector operations can be readily accelerated by photons due to the natural parallelism of bosons[9]. In the past decades, various photonic computing architectures have been demonstrated to perform vector-matrix operation in optical domain, *i.e.* Stanford structure[10-12], Reck scheme[13-16], deep diffraction architecture[17-20], micro-ring resonator (MRR) array[21-26], *etc.* All these approaches can only perform limited scalability of the linear vector operation due to two fundamental issues. As the optical matrix is adopted to perform linear transformation of the input vector encoded on light beam, the basic units in the computing architecture, *i.e.* liquid cells, beam splitters, meta-atoms, *etc.*, would be tightly interconnected or highly coupled due to the unavoidable light beam interaction. Thus, high-dimensional optical vector-matrix operations cannot be achieved by simply duplicating the basic unit. Moreover, another issue is how to fuse the linear optics with the existing electronic digital computing systems. Due to significant obstacles in implementing all-optical computers[27], optoelectronic hybrid computing architecture may be the most promising solution[28] by performing the linear vector operations in the optical domain and nonlinear operations in the electronic domain. This requires a large number of analog-to-digital converters (ADCs) and digital-to-analog converters (DACs) for exchanging analog data in optical domain and digital data in electronic domain. In fact, such conversion between digital and analog data is the main limiting factor to significantly scale up the computing dimensionality[29]. In addition, the slow speed of ADC and DAC remains the main obstacle to high-speed operation with linear optical computing. To address this issue, an all-analog chip combining electronic and light computing (ACCEL)[30] was presented, but it can only serve as a specific processor for the vision task. Therefore, a universal optoelectronic hybrid computing architecture, which does not rely on light beam interaction and can allow high-speed conversions between analog and digital data is urgently needed to address the two fundamental issues so that flexible scalability can be achieved to alleviate the heavy computing demands of future AI technology.

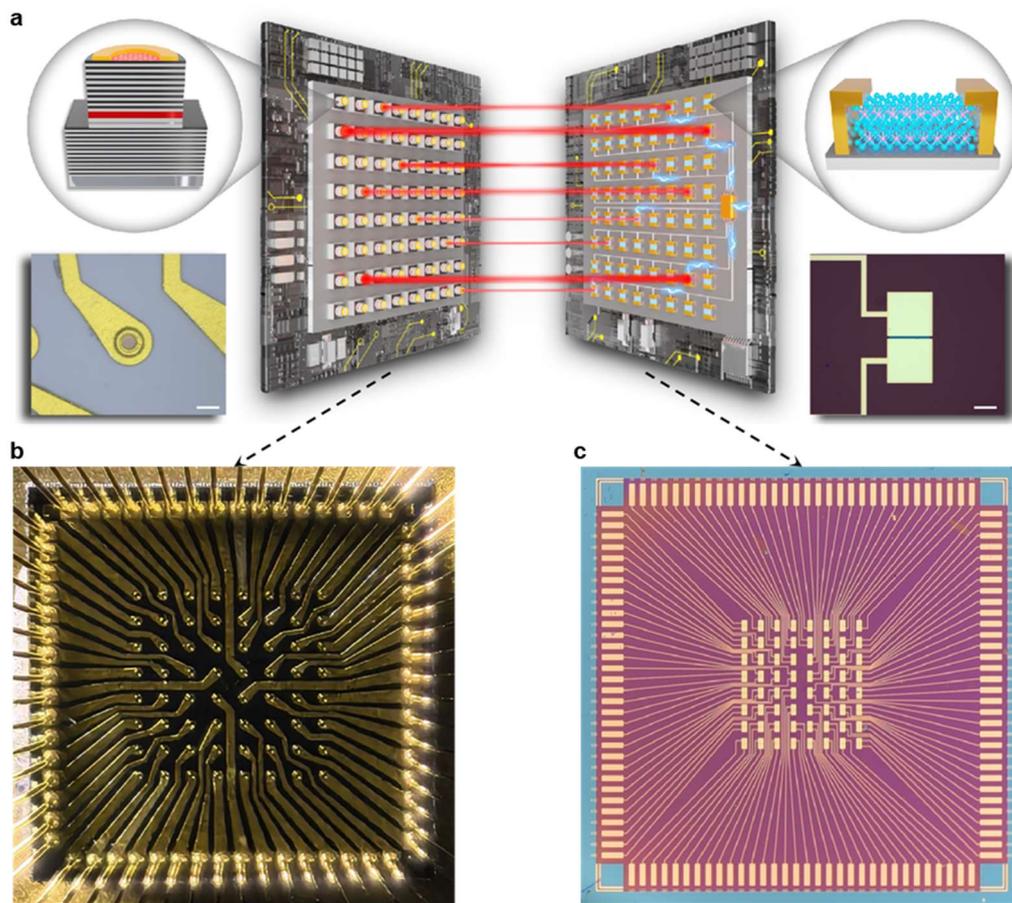

**Fig.1 | Architecture of SUANPAN. a,** The schematic diagram of SUANPAN architecture, consisting of a light-emitter array, a PD array and some necessary electronic hardware. Left inset shows the schematic and microscope photograph of a single VCSEL. Scale bar is 20 μm. Right inset shows the schematic and microscope photograph of a single MoTe$_2$ PD. Scale bar is 100 μm. **b,** The optical image of the VCSEL array. **c,** The optical image of the MoTe$_2$ PD array.

Inspired by the traditional Chinese Suanpan or abacus, in which the mathematical operations are carried out by moving the beads, we propose and demonstrate a photonic SUANPAN architecture to pursue a brand-new analog-digital hybrid computing paradigm. The basic computing unit in SUANPAN consists of a pair of light-emitter and photodetector (PD), and can achieve Bit Encoding and Analog Detecting, denoted as BEAD. Our proposed SUANPAN is formed by BEAD array, and thus can easily scale up by duplicating the independent BEAD. Instead of optical matrix operations in the current popular optical computing architectures, SUANPAN is designed only for the optical inner product of two vectors. The element values of one vector are encoded on the output intensity of light-emitter and realized by controlling the duty ratio of driving current. While the element values of the other vector are encoded by the position of *M* BEADs to achieve *M*-bit quantization. The photocurrent of the PD would be proportional to the multiplication of the light intensity and photoresponsivity, and the final result of the inner product can be obtained by the summation of all the photocurrents. There is no DAC and only one ADC required in SUANPAN architecture, so it can be readily merged with existing electronic digital computing system. As a proof of principle, the SUANPAN architecture is implemented using an 8×8 vertical cavity surface emission laser (VCSEL) array and an 8×8 MoTe$_2$ two-dimensional (2D) material PD array. The experimental results of vector inner product operations show that the calculation fidelity can be as high as >98% for various bit precisions (1-bit, 2-bit, 4-bit and 8-bit), and >95% for various vector dimensionalities (@4-bit precision), respectively. Furthermore, such SUANPAN with 64 BEADs has been successfully reconfigured to perform two typical AI tasks, Ising machine and ANN. 1024-dimensional randomly generated Ising problem is successfully solved, which is the highest dimensional optical Ising machine with heuristic algorithm. Meanwhile, a competitive classification accuracy of 84%~88% is achieved for MNIST handwritten digit dataset by properly setting SUANPAN for an ANN. Since there is no interaction among the propagating light beams of all BEADs and only the output currents of all BEADs are connected, SUANPAN architecture could be extremely scalable by increasing the number of BEADs with no additional loss or error as well as flexibly reconfigurable and programmable by properly arranging the BEADs for different computational tasks. We believe that our proposed photonic SUANPAN is capable to serve as a fundamental linear vector machine and is potential to enhance the computing power for future various AI applications.

**SUANPAN architecture**

The proposed SUANPAN architecture consists of a light-emitter array and a PD array as well as some necessary electronic hardware as schematically shown in Fig.1a. Here, a VCSEL array and a $MoTe_2$ 2D material PD array are fabricated to demonstrate the SUANPAN architecture as shown in Fig.1b and Fig.1c, respectively. The schematic diagram and microscope photographs of a single VCSEL and $MoTe_2$ PD are also shown in the insets of Fig.1a. SUANPAN architecture is designed for vector inner product, also known as multiply-accumulate (MAC) operation. Firstly, for multiply operation, each PD is well aligned with a corresponding light-emitter, therefore the photocurrent of the PD would be proportional to the multiplication of the light intensity and photoresponsivity due to the linear optical response[31]. Then, for add operation, the photocurrent of each PD is connected together, therefore the output current would be the sum of all PDs due to the Kirchhoff's law. In this way, the multiply-accumulate operation is naturally performed through the emission and detection in SUANPAN architecture. In addition, one more important issue is how to encode the vectors on to the light-emitter array and PD array. In the traditional Chinese Suanpan, numbers are represented by the different positions of beads, and the mathematic operations are carried out by moving beads. Inspired by that, the basic computing unit in SUANPAN consists of a pair of light-emitter and PD, and can achieve Bit Encoding and Analog Detecting, named as BEAD. The vector encoding and operating would rely on controlling the on-off state of BEADs.

For deep insight, a simple example of one BEAD is considered. As shown in Fig.2a, the multiplier *a* is encoded in the intensity of light-emitter by controlling the duty ratio of driving current, which is done by a digital counter according to the clock cycles without DAC. The bit precision depends on the time-slot numbers of splitting the period. With the constant period, the more time-slots there are, the more quantization bits for *a* can be achieved. In this work, the period of light emission is divided into 100 slots, so that $a$ can be taken as 0, 1, …, 100. The multiplier $b$ is encoded on the on-off state of the BEAD, by turning on (green arrow) or off (gray arrow) the light-emitter, respectively. Hence, there are two states to encode $b$, $b=0$ or $b=1$, known as 1-bit quantization. The photocurrent would be proportional to $a \times 0$ or $a \times 1$, which is the result of $a \times b$, $b=0,1$. For more bit quantization of $b$, more BEADs are employed and form a set of them. Considering 2-bit quantization of $b$, there would be two BEADs in one set as shown in the white box of Fig.2b. The method of encoding $a$ is also as aforementioned, while encoding $b$ should employ two BEADs simultaneously and thus form four combinations of on-off states, which is corresponding to the four values of 2-bit quantization. From the perspective of binary encoding, such two BEADs are equivalent to the two bits in the binary representation of $b$, while the position of PDs is considered as the bit-position. For example, if $b=2$, then the binary representation would be $b=\overline{10}$, which means the first BEAD is at off-state and the second one is at on-state. As mentioned above, this is quite similar to the Chinese traditional Suanpan, which represents numbers according to the position of beads and carries out mathematical operations by moving beads up and down. Different bits represent different weights in binary representation, and this can be achieved by setting the photoresponsivity of these two PDs as $2^0$ to $2^1$. To properly manipulate the photoresponsivity, we design and fabricate 2D material photoconductive detectors (more details will be discussed in Result section). By combining the photocurrents of these two PDs together, the output result is $a \times b$ as shown in Fig.2b. Further, in this way, $M$-bit quantization of $b$ can be achieved with a set of $M$ BEADs, so that SUANPAN can encode the range of $b$ from 0 to $2^M-1$ and achieve the multiplication of $a$ and $b$, as shown in Fig.2c.

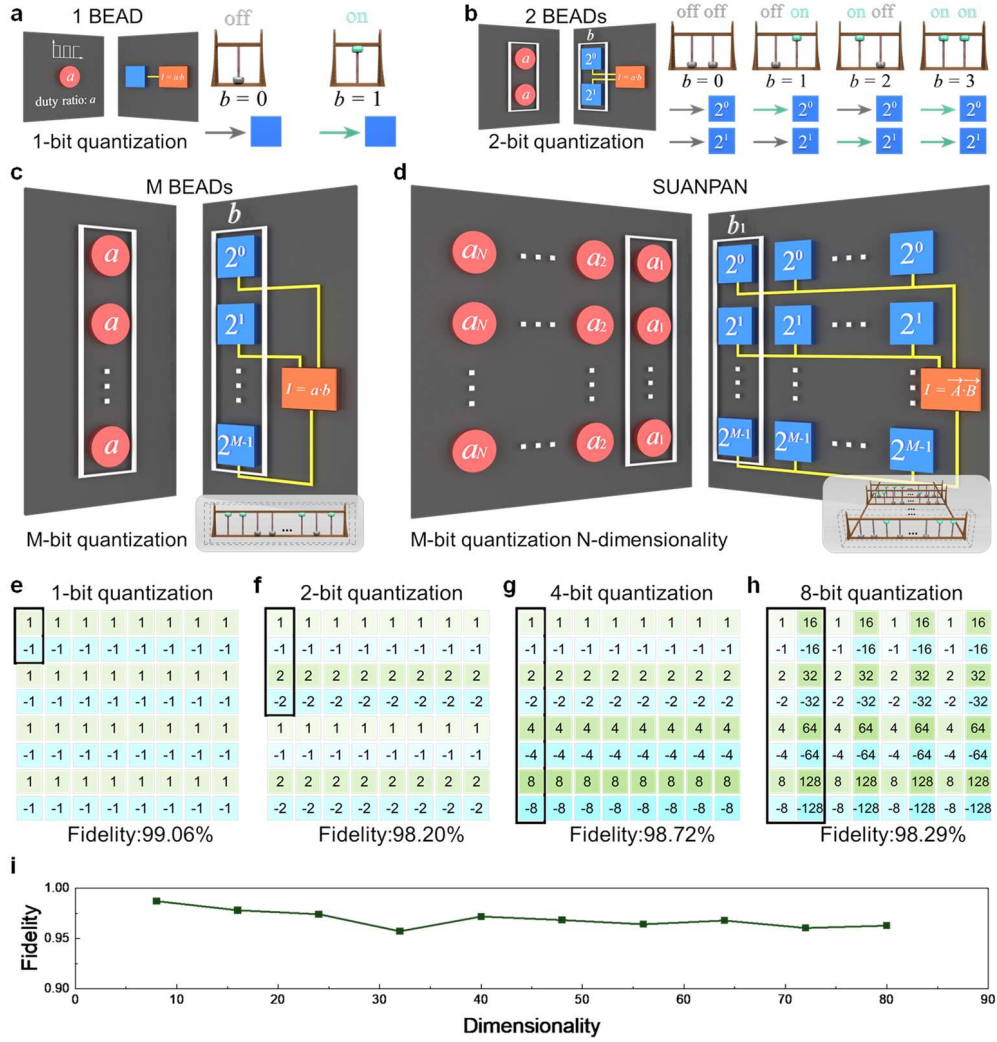

**Fig.2 | Operation principle of SUANPAN. a,** The mechanism of performing $a \times b$, where $b$ is 1-bit quantization with one BEAD. Inset: the on-off state of 1 BEAD in SUANPAN comparing with 1 bead in Chinese Suanpan. **b,** The mechanism of performing $a \times b$, where $b$ is 2-bit quantization with two BEADs. Inset: the on-off states of 2 BEADs in SUANPAN comparing with 2 beads in Chinese Suanpan. **c,** The mechanism of performing $a \times b$, where $b$ is $M$-bit quantization with $M$ BEADs. The white box represents a set of $M$ BEADs. Inset: $M$ beads in Chinese Suanpan as a comparison. **d,** The mechanism of performing vector inner product $A \cdot B$ with $N$ sets of $M$ BEADs. Inset: $N$ sets of Chinese Suanpan as a comparison. **e-h,** The configuration of SUANPAN and experimental fidelity for 1-bit, 2-bit, 4bit and 8-bit quantization, respectively. **i,** The experimental fidelity for 4-bit quantization versus the dimensionality scales.

As mentioned above, $a$ is digitally encoded to the duty ratio of light emission and $b$ is digitally encoded to the on-off states of the BEADs in one set, while the output result is the analog photocurrent from the PDs. Thus, the basic unit, consisting of a pair of light-emitter and PD, is actually operated as Bit Encoding and Analog Detecting, which is thus abbreviated as BEAD. Since the number of $a$ and $b$ are encoded in time and space domain respectively, SUANPAN architecture can perform the multiplication of any desired bit precision theoretically. The vector inner product can be further achieved by employing $N$ sets of BEADs and connecting their output photocurrents together (Fig.2d). It should be mentioned that SUANPAN architecture can also handle negative numbers by applying reversed bias voltage of the corresponding PD. Considering both positive/negative numbers, $2 \times M$ BEADs should be utilized for each set and the details are provided in Extended Data Fig.1. Through this encoding method, $M$-bit quantization and $N$-dimensional signed vector inner product can be performed by SUANPAN with $N \times 2 \times M$ BEADs at a time while each element of the input vectors can be flexibly programmed. Furthermore, the number of utilized sets ($N$) as well as the number of BEADs in each set ($M$) can be reconfigurable according to specific calculation requirements in terms of the dimensionality and bit precision. At the output, the photocurrents of all PDs are connected together, so that only one ADC is required to transform the total analog current into digital signals. Also, since only 1-bit information would be encoded on a single BEAD, the properly settled but fixed bias voltage would be applied on each BEAD. Thus, there are no requirement for ADC array or DAC array in SUANPAN, which is actually the main limiting factor for the hybrid computing with both digital and analog operations. Last but not least, since there is no interaction among the propagating light beams of all BEADs, SUANPAN architecture can be extremely scalable by increasing the number of independent BEADs with no additional loss or error. On one hand, the number of utilized BEADs can be easily increased by integrating more components on one pair of light-emitter array and PD array chips. On the other hand, distributed computing can also be achieved by simply connecting multiple pairs of chips together to scale up the computing power more. Therefore, SUANPAN is a programmable, reconfigurable and scalable computing architecture, which is capable to serve as a general vector inner product accelerator for the existing computing system.

**Result**

To demonstrate the SUANPAN architecture, a pair of VCSEL and MoTe$_2$ PD are employed to form the BEAD. As a light-emitter, VCSEL can readily achieve high-speed modulation as well as large-scale array. For PD, 2D material of MoTe$_2$ is utilized for three reasons: (1) The photoresponsivity of 2D material PD can be flexibly controlled by the bias voltage. (2) The high carrier mobility in 2D material can support high-speed detection, which is an important issue for high-speed computing. (3) 2D material can be heterogeneously integrated with other material platform. Therefore, 2D material PD is potentially integrated with light-emitter array in the future, which will be explained in detail in the Discussion section. Thus, we have fabricated both the VCSEL and MoTe$_2$ PD array chip with 8×8 components, respectively. The fabrication process can be found in Methods and Extended Data Fig.2, and the characteristics of each VCSEL and PD can be found in Extended Data Fig.3-5. Our fabricated VCSELs and PDs show good uniformity and stability, which are very significant in the optical computing architecture. To align each pair of VCSEL and PD, an optical imaging system is built up (refer to Extended Data Fig.7). Such an imaging system would be not necessary by integrating the 2D material PD array with light-emitter array on a single chip in the future. To verify the functionality of SUANPAN architecture, random vector inner products are performed with various bit precisions and dimensionalities. Finally, two typical AI tasks of the Ising machine and ANN, are performed to present the ability of SUANPAN architecture.

The inner product of random vectors is the foundation for different AI tasks and the signed vector inner product with precision of 1-bit, 2-bit, 4-bit and 8-bit are demonstrated. For 1-bit precision signed vector, 2 BEADs are required in each set, and there are 32 sets with 64 BEADs. Therefore, 32-dimensional vector inner product can be done at one time. Similarly, for 2-bit, 4-bit and 8-bit quantization, the corresponding dimensionality would be 16, 8 and 4 respectively. To achieve higher dimensional vector inner product, time multiplexing can also be employed, which is according to accumulating multiple times of calculations. For each bit precision, the configuration of SUANPAN would be properly settled and the corresponding bias voltage of each PD is shown in Fig.2e-h. For each bit precision, 1000 pairs of signed vectors are randomly generated and performed by SUANPAN. The calculation accuracy is evaluated by the fidelity expressed as:

$$\text{Fidelity}(x, y) = \frac{x^T y}{\|x\| \cdot \|y\|} \quad (1)$$

Here, 1000 true values are calculated in a computer and denoted as a vector *x*, and the calculated results in SUANPAN are recorded as a vector *y*. As shown in equation (1), the fidelity describes the parallel degree between *x* and *y*, while the scaling error can be excluded. The statistical results are shown in Fig.2e-h and all of those fidelities are higher than 98%. These results indicate that SUANPAN can perform accurate calculation of signed vector inner products with different bit precisions. Specifically, the fidelity of 4-bit precision signed vector inner product with different dimensionalities are shown in Fig.2i. As the dimensionality increases, the computational fidelity remains above 95%, which also demonstrates the scalability of SUANPAN architecture. Due to the high fidelity in executing random signed vector inner products with different bit precisions and dimensionalities, SUANPAN architecture can be flexibly utilized to further demonstrate more specific computing tasks. In next sections, both the Ising machine and ANN would be considered.

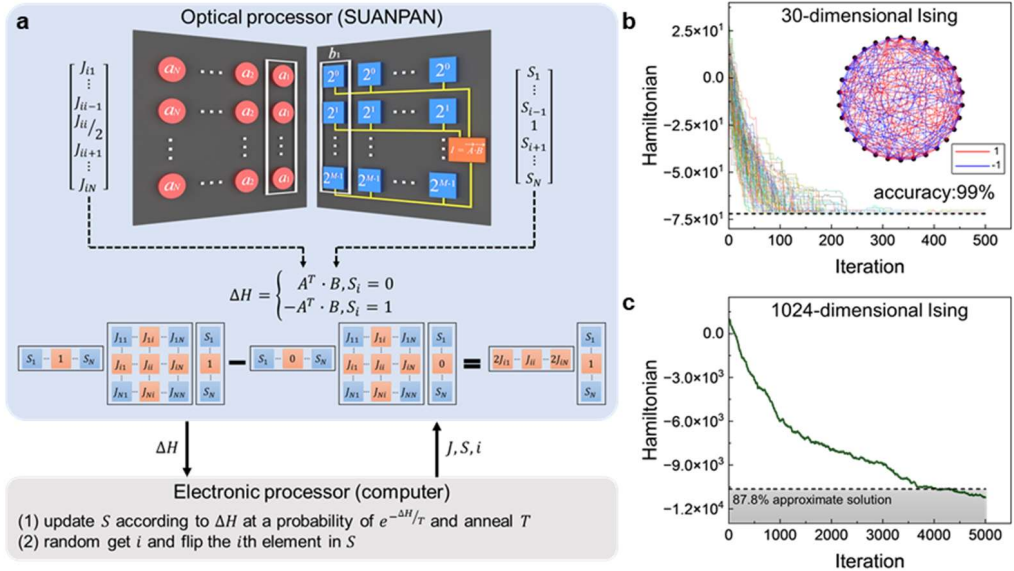

**Fig.3 | Photonic Ising machine with SUANPAN architecture. a,** The system architecture of Ising machine utilizing SUANPAN and a digital computer as optoelectronic hybrid computing. Blue box: the vector inner product performed by SUANPAN, which is equivalent to calculate $\Delta H$. Gray box: nonlinear operations performed by the digital computer. **b,** 100 experimental annealing curves of a random 30-dimensional Ising problem. Dashed line: ground state. Inset: the random 30-dimensional Ising model. **c,** Experimental annealing curve of a random 1024-dimensional Ising problem. Dashed line: 87.8% approximate solution.

For the decision-making task in AI applications, the combinatorial optimization problems are ubiquitous and usually non-deterministic polynomial (NP). One approach to process such NP-hard problems is mapping them to Ising problem[32], which is a typical combinatorial optimization problem and also known as quadratic unconstrained binary optimization (QUBO) problem. An $N$-dimensional Ising problem can be defined by an interaction matrix $J$, which is a symmetric matrix of $N \times N$ dimensionality. For a given interaction matrix $J$, the Hamiltonian of Ising problem is defined as follows:

$$H = S^T J S \qquad (2)$$

Solving Ising problem is to find the specific vector $S$ that minimizes the Hamiltonian, which is denoted as the ground state. Since the elements in $S$ can only take 0 or 1, the dimensionality of the solution space is $2^N$ for an $N$-dimensional Ising problem. With the development of AI, various heuristic algorithms have been developed to solve the Ising problem and the so-called Ising machines have been demonstrated on various computing platforms. Among them, the simulated annealing (SA) algorithm[33] is combined with optical computing platforms to form a photonic Ising machine due to the parallelism of light beam. Here, SUANPAN is set to serve as a hardware solver for Ising problem. The details of SA algorithm are shown in Ref.[33], which consists of initialization, $n$ iterations and output. In each iteration, one random element of $S$ is flipped and then the variation of Hamiltonian $\Delta H$ is calculated. After that, the vector of $S$ would be accepted or not according to $\Delta H$. Obviously, the Hamiltonian should be calculated in each iteration, which is actually the main computational burden in SA algorithm with $O(N^2)$ complexity. Through some identity transformations, $\Delta H$ can be transformed into an $N$-dimensional vector inner product, which can be readily performed by SUANPAN as shown in Fig.3a. Therefore, there are mainly two functional units to build up such an Ising machine here. The first one is an electronic processor (gray box in Fig.3a), to execute the updating and flipping operations. The other is SUANPAN (yellow box in Fig.3a), where vector inner product would be performed to calculate $\Delta H$. Here, as shown in Fig.3a, the first vector is the $i$th row of matrix $J$ with some proper modifications. Since $J$ is a continuous variable matrix, it is encoded on light-emitter array to achieve a high bit precision. The second vector is according to vector $S$, in which the element only takes 0/1. Thus, it is encoded on PD array with 1-bit precision. It should be mentioned that if $S_i$ is flipped from 0 to 1, then the result of the vector inner product would be $\Delta H$. Otherwise, if $S_i$ is flipped from 1 to 0, then negative sign should be taken.

We first solve a 30-dimensional randomly generated Ising problem by SUANPAN, which is the highest dimensionality of arbitrarily connected Ising problem reported previously in a programmable photonic Ising machine[34]. The current SUANPAN with 64 BEADs can calculate 32-dimensional vector inner product at one time with 1-bit precision, and thus it can be employed to solve 30-dimensional Ising problems without time-division multiplexing. The iteration number is set as 500 and the solving process is repeated with 100 rounds while the full parameters of the SA algorithm are provided in Extended Data Table1. Each curve in Fig.3b represents the Hamiltonian evolution with the increasing of iterations during each solving process, and 99 curves eventually converged to the lowest ground state. The Hamiltonian decreases very rapidly in the initial 100 iterations. After about 230 iterations, most of the curves have already been very close to the ground state. Finally, a accuracy of 99% is achieved by SUANPAN, which is much higher than the existing optical programmable Ising machine[34].

To further validate the scalability of SUANPAN architecture, a randomly generated 1024-dimensional Ising problem is considered, in which dimensionality is more than an order of magnitude comparing with the previously reported programmable photonic Ising machine based on heuristic algorithm. Here, the required 1024-dimensional vector inner product is decomposed into 32-dimensional one with time division multiplexing for 32 times. The iteration number is set as 5000 and full parameters of the SA algorithm are provided in Extended Data Table1. For high-dimensional Ising problem, it is hard to obtain the true Hamiltonian value of the ground state, and thus an approximate solution with 87.8% of the ground state is usually set as a criteria for successful solution[35-37], as dashed line shown in Fig.3c. Such 1024-dimensional Ising problem is successfully solved by SUANPAN as the annealing curve fall below the criteria after ~4000 iterations, as shown in Fig.3c. The high convergence rate and high dimensionality in solving Ising problem can fully validate the programmability, reconfigurability and computational stability of SUANPAN architecture, which is capable to serve as a programmable Ising machine.

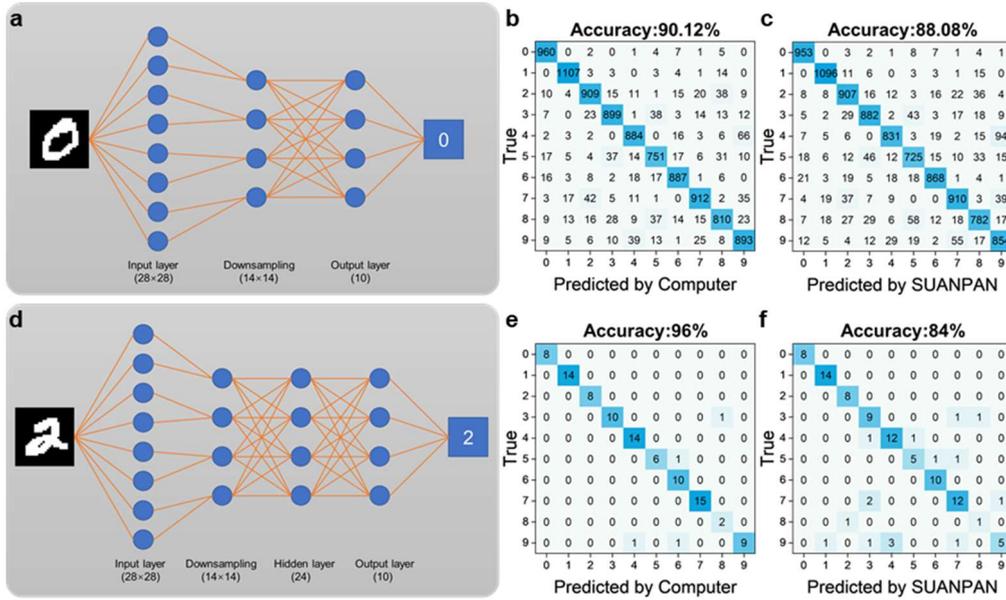

**Fig.4 | Artificial neural network with SUANPAN architecture. a,** The pre-trained one-layer ANN for MNIST dataset. **b-c,** The accuracy and confusion matrix of the one-layer ANN performed by computer and SUANPAN, respectively. **d,** The pre-trained double-layer ANN for MNIST dataset. **e-f,** The accuracy and confusion matrix of the double-layer ANN performed by computer and SUANPAN, respectively.

ANN is another typical task in modern AI and various physical neural networks (PNNs) have been applied for the tasks of visual perception, speech recognition, subject classification, *etc*. Among them, optical neural networks are quite promising also due to the parallelism of light beam[4,5]. In PNNs, silico training is usually required to avoid errors caused by differences between simulation and practical devices. Unlike that, SUANPAN architecture can directly map pre-trained ANNs, in which the input data needs to go through a linear part and a nonlinear part during each layer. In the linear part of ANN, the vector matrix multiplication is calculated between input data and weight matrix, while in the nonlinear part, a nonlinear activation function is performed. Correspondingly, the linear matrix multiplication, which can be considered as a set of vector inner product, would be performed by SUANPAN, and the nonlinear function would be calculated by electronic processor. Therefore, through time-division multiplexing, SUANPAN architecture can execute ANNs of arbitrary depth and arbitrary number of nodes in theory. As proof of principle, both single and double layer ANNs are performed with SUANPAN architecture as shown in Fig.4a and 4d, respectively. Here, MNIST handwritten digit is utilized as dataset, and stochastic gradient descent[38] (SGD) is utilized as training method. The nonlinear function of the output layer is Softmax function, and the nonlinear function of the hidden layer is Relu function. The weights of the single-layer ANN and double-layer ANN are 4-bit precision and 6-bit precision according to simulations, respectively (the details shown in Extended Data Fig.8). For the inference result of single-layer ANN, the confusion matrix of 10000 pictures in test dataset calculated by computer and SUANPAN are shown in Fig.4b and 4c, respectively. The approaching classification accuracies are 88.08% and 90.12% for SUANPAN and computer, respectively. It can be seen that the results processed by SUANPAN are quite consistent with those processed by a computer, and only a little deterioration is introduced. These results indicate a quite high computing accuracy in SUANPAN architecture. It should be mentioned that the classification accuracy is not limited by the SUANPAN, but by the pre-trained model itself, which could be further improved through optimizing the network model and the training method. While for double-layer, only the first 100 pictures in MNIST test dataset are performed as a proof of principle verification. The confusion matrix and accuracy calculated by computer and SUANPAN are shown in Fig.4e and 4f. It can be noticed that the classification accuracy of double-layer ANN calculated by computer is much higher than one-layer, while that calculated by SUANPAN is lower than one-layer. The reason might be the device performance of $MoTe_2$ PD array has deteriorated after two months of testing. Anyway, we believe that the above results of ANN can still validate the feasibility of SUANPAN architecture.

**Discussion**

In this work, we have proposed and demonstrated the photonic SUANPAN architecture to perform the vector inner product operations. Instead of the reported optical computing architecture maximizing the interaction of light beams, we utilize independent pairs of light-emitter and PD to form a programmable, reconfigurable and scalable computing architecture that can be compatible with the existing computing system. Our fabricated SUANPAN with 64 pairs of VCSEL and $MoTe_2$ PD shows high computing fidelities for randomly generated vector inner products and demonstrates two typical AI tasks of the Ising machine and ANN. For the Ising problem, 1024-dimensional problem is successfully solved, which is the highest dimensional optical Ising machine with heuristic algorithm. For ANN, a competitive classification accuracy of 84%~88% is achieved for MNIST handwritten digit dataset. There are two main contributions in this work, the SUANPAN computing architecture and the Bit Encoding and Analog Detecting computing paradigm, which actually hold together.

Firstly, for the SUANPAN architecture, it breaks through the traditional mindset of obtaining optical matrix transformations through interaction of light beams. Instead, there is no interaction among those propagating light beams of all BEADs. Therefore, the SUANPAN architecture can be decomposed into BEADs as independent computing units. The scalability, reconfigurability and programmability of the SUANPAN architecture are only based on the duplication, recombination and modulation of BEAD without any additional cost. Compared with other optical matrix transformations through interaction between light beams, SUANPAN architecture possesses these following advantages: (1) With massive and industrial replication of BEADs, SUANPAN architecture can theoretically be infinitely scalable. (2) SUANPAN architecture can be flexibly reconfigured and programmed to perform various specific computing tasks. (3) The correction of SUANPAN architecture only considers the intensity of light beam, and there is no requirement to correct the phase of the light beams for interaction. (4) Even if one BEAD is damaged during fabrication or operation, it does not affect other BEADs, and the only result is decreasing the operating dimensionality. In short, the core idea of SUANPAN architecture is to decompose the whole computing architecture into simple and independent units, since no interaction among each unit would bring more scalability and programmability to the whole optical computing architecture.

Secondly, for the Bit Encoding and Analog Detecting computing paradigm, it provides a promising solution for optoelectronic analog-digital hybrid computing. For traditional DA conversion in optoelectronic computing, the digital electronic signal is converted to analog optical signal in each single device. Therefore, each single device is required a DAC. However, in Bit Encoding and Analog Detecting paradigm, the $M$-bit digital electronic signal is converted to analog with in a set of $M$ BEADs, where each BEAD only represents 1-bit information. Therefore, no DAC is required in such computing paradigm. At the same time, only one ADC is required to convert the sum up photocurrent into electronic digital signal as the result. Therefore, the Bit Encoding and Analog Detecting computing paradigm greatly reduces the burden introduced by ADC and DAC. In fact, it is also an important issue for the arbitrary scalability of SUANPAN architecture. Additionally, this computing paradigm can be extended beyond SUANPAN architecture and introduced into other computing architectures to reduce the obstructive effects of ADC and DAC.

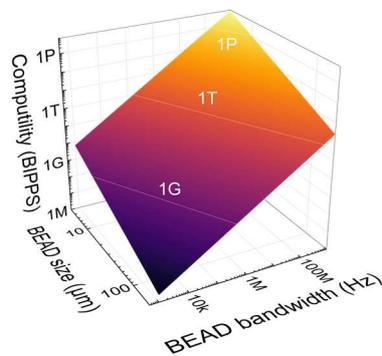

**Fig.5 | The estimated computility evolution with decreasing BEAD size and increasing BEAD bandwidth.**

Finally, our work is a preliminary verification of the feasibility of SUANPAN architecture, and there is still a lot of room for improvement within the specific implementation. It is obvious that the light-emitter array and the 2D material PD array can be integrated into a single chip due to the heterogeneous integration of 2D materials. In this way, the imaging system is not required to align the light-emitter array and PD array, and further 3D stacking integration can be achieved to further expand the scalability. In addition, a detailed analysis is provided here for the future development of the computility. In order to evaluate the computility of SUANPAN architecture, a metric is defined as BIPPS here, which is bits of inner product per second. Compared to floating point operations per second (FLOPS), if the computational task is $M$-bit quantized, then $M$ BIPPS is equivalent to 1 FLOPS. For SUANPAN architecture, computility is determined by two factors, the BEAD size and bandwidth. Since BEAD consists of a pair of light-emitter and PD, the BEAD size is defined as the larger one between light-emitter and PD, while the BEAD bandwidth is defined as the smaller one between the modulating bandwidth of light-emitter and the response bandwidth of PD. Considering within the utilized chip area of 1 cm$^2$, the computility is obtained as follow:

$$\text{Computility} = \left(\frac{1\text{ cm}}{\text{BEAD size}}\right)^2 \cdot \text{BEAD bandwidth} \qquad (3)$$

where the first term equals the number of BEADs on a chip area of 1 cm$^2$. Therefore, the computility can be increased by reducing the BEAD size and increasing the BEAD bandwidth as shown in Fig.5. Due to the research on various high-speed nano-lasers[39-41] and nano-detectors[42,43], the computility could achieve over 1P BIPPS/cm$^2$ for <10 μm BEAD size and >1 GHz BEAD bandwidth. Therefore, SUANPAN architecture is capable as an attractive and practicable photonic linear vector machine in the foreseeable future

1       Cetinic, E. & She, J. Understanding and Creating Art with AI: Review and Outlook. *ACM Transactions on Multimedia Computing, Communications, and Applications* **18**, 1-22. (2022).

2       Rajpurkar, P. *et al.* AI in health and medicine. *Nat. Med.* **28**, 31-38. (2022).

3       LeCun, Y. *et al.* Deep learning. *Nature* **521**, 436-444. (2015).

4       Wetzstein, G. *et al.* Inference in artificial intelligence with deep optics and photonics. *Nature* **588**, 39-47. (2020).

5       Fu, T. *et al.* Optical neural networks: progress and challenges. *Light Sci Appl* **13**, 263. (2024).

6       Mohseni, N. *et al.* Ising machines as hardware solvers of combinatorial optimization problems. *Nature Reviews Physics* **4**, 363-379. (2022).

7       Laydevant, J. *et al.* Training an Ising machine with equilibrium propagation. *Nat Commun* **15**, 3671. (2024).

8       Nikhar, S. *et al.* All-to-all reconfigurability with sparse and higher-order Ising machines. *Nat Commun* **15**, 8977. (2024).

9       Zhou, H. *et al.* Photonic matrix multiplication lights up photonic accelerator and beyond. *Light Sci Appl* **11**, 30. (2022).

10      Goodman, J. W. *et al.* Fully parallel, high-speed incoherent optical method for performing discrete Fourier transforms. *Opt Lett* **2**, 1-3. (1978).

11      Spall, J. *et al.* Fully reconfigurable coherent optical vector-matrix multiplication. *Opt Lett* **45**, 5752-5755. (2020).

12      Wang, T. *et al.* An optical neural network using less than 1 photon per multiplication. *Nat Commun* **13**, 123. (2022).

13      Reck, M. *et al.* Experimental realization of any discrete unitary operator. *Phys Rev Lett* **73**, 58-61. (1994).

14      Shen, Y. *et al.* Deep learning with coherent nanophotonic circuits. *Nature Photonics* **11**, 441-446. (2017).

15      Roques-Carmes, C. *et al.* Heuristic recurrent algorithms for photonic Ising machines. *Nat Commun* **11**, 249. (2020).

16      Pai, S. *et al.* Experimentally realized in situ backpropagation for deep learning in photonic neural networks. *Science* **380**, 398-404. (2023).

17      Lin, X. *et al.* All-optical machine learning using diffractive deep neural networks. *Science* **361**, 1004-1008. (2018).

18      Yan, T. *et al.* Fourier-space Diffractive Deep Neural Network. *Phys Rev Lett* **123**, 023901. (2019).

19      Zhou, T. *et al.* Large-scale neuromorphic optoelectronic computing with a reconfigurable diffractive processing unit. *Nature Photonics* **15**, 367-373. (2021).

20      Fu, T. *et al.* Photonic machine learning with on-chip diffractive optics. *Nat Commun* **14**, 70. (2023).

21      Tait, A. N. *et al.* Broadcast and Weight: An Integrated Network For Scalable Photonic Spike Processing. *Journal of Lightwave Technology* **32**, 4029-4041. (2014).

| | |
|---|---|
| 22 | Deng, Y. & Chu, D. Coherence properties of different light sources and their effect on the image sharpness and speckle of holographic displays. *Sci Rep* **7**, 5893. (2017). |
| 23 | Feldmann, J. *et al.* All-optical spiking neurosynaptic networks with self-learning capabilities. *Nature* **569**, 208-214. (2019). |
| 24 | Feldmann, J. *et al.* Parallel convolutional processing using an integrated photonic tensor core. *Nature* **589**, 52-58. (2021). |
| 25 | Bai, B. *et al.* Microcomb-based integrated photonic processing unit. *Nat Commun* **14**, 66. (2023). |
| 26 | Dong, B. *et al.* Partial coherence enhances parallelized photonic computing. *Nature* **632**, 55-62. (2024). |
| 27 | Kazanskiy, N. L. *et al.* Optical Computing: Status and Perspectives. *Nanomaterials* **12**. (2022). |
| 28 | Dan, Y. *et al.* Optoelectronic integrated circuits for analog optical computing: Development and challenge. *Frontiers in Physics* **10**. (2022). |
| 29 | Kim, S. *et al.* Neuro-CIM: ADC-Less Neuromorphic Computing-in-Memory Processor With Operation Gating/Stopping and Digital–Analog Networks. *IEEE Journal of Solid-State Circuits* **58**, 2931-2945. (2023). |
| 30 | Chen, Y. *et al.* All-analog photoelectronic chip for high-speed vision tasks. *Nature* **623**, 48-57. (2023). |
| 31 | Huo, N. & Konstantatos, G. Recent Progress and Future Prospects of 2D‐Based Photodetectors. *Advanced Materials* **30**. (2018). |
| 32 | Lucas, A. Ising formulations of many NP problems. *Frontiers in Physics* **2**. (2014). |
| 33 | Van Laarhoven, P. J. & Aarts, E. H. *Simulated annealing: theory and application*. (Springer, 1987). |
| 34 | Ouyang, J. *et al.* On-demand photonic Ising machine with simplified Hamiltonian calculation by phase encoding and intensity detection. *Communications Physics* **7**. (2024). |
| 35 | Yamamoto, Y. *et al.* Coherent Ising machines—optical neural networks operating at the quantum limit. *npj Quantum Information* **3**. (2017). |
| 36 | Haribara, Y. *et al.* Computational Principle and Performance Evaluation of Coherent Ising Machine Based on Degenerate Optical Parametric Oscillator Network. *Entropy* **18**. (2016). |
| 37 | Goemans, M. X. & Williamson, D. P. Improved approximation algorithms for maximum cut and satisfiability problems using semidefinite programming. *Journal of the ACM* **42**, 1115-1145. (1995). |
| 38 | Ketkar, N. *Deep learning with Python*. (Springer, 2017). |
| 39 | Jeong, K. Y. *et al.* Recent Progress in Nanolaser Technology. *Advanced Materials* **32**. (2020). |
| 40 | Du, W. *et al.* Nanolasers Based on 2D Materials. *Laser & Photonics Reviews* **14**. (2020). |
| 41 | Zhang, Q. *et al.* Halide Perovskite Semiconductor Lasers: Materials, Cavity Design, and Low Threshold. *Nano Lett* **21**, 1903-1914. (2021). |
| 42 | Long, M. *et al.* Progress, Challenges, and Opportunities for 2D Material Based Photodetectors. *Advanced Functional Materials* **29**. (2018). |
| 43 | Liu, C. *et al.* Silicon/2D-material photodetectors: from near-infrared to mid-infrared. *Light: Science & Applications* **10**. (2021). |
| 44 | Xu, X. *et al.* Millimeter-Scale Single-Crystalline Semiconducting MoTe(2) via Solid-to-Solid Phase Transformation. *J Am Chem Soc* **141**, 2128-2134. (2019). |
| 45 | Pan, Y. *et al.* Heteroepitaxy of semiconducting 2H-MoTe2 thin films on arbitrary surfaces for large-scale heterogeneous integration. *Nature Synthesis* **1**, 701-708. (2022). |